\documentclass{article}
\usepackage[T1]{fontenc} 
\usepackage[utf8]{inputenc} 
\usepackage{ismir,amsmath,cite,url}
\usepackage{graphicx}
\usepackage{color}

\usepackage{lineno}

\usepackage{soul}
\usepackage{color}
\usepackage{amssymb}
\usepackage{enumitem}

\usepackage{booktabs}
\usepackage{multirow}
\usepackage{url}
\usepackage[table]{xcolor}

\setlist{topsep=4pt, itemsep=0pt, leftmargin=*}

\title{Efficient Supervised Training of Audio Transformers for Music Representation Learning}

\multauthor
{Pablo Alonso-Jim\'enez \hspace{1cm} Xavier Serra \hspace{1cm} Dmitry Bogdanov} {
Music Technology Group, Universitat Pompeu Fabra, Barcelona \\
{\tt\small pablo.alonso@upf.edu}
}

\def\authorname{P. Alonso-Jim\'enez, X. Serra, and D. Bogdanov}

\usepackage[bookmarks=false,pdfauthor={\authorname},pdfsubject={\papersubject},hidelinks]{hyperref}

\begin{document}

\maketitle
\begin{abstract}
In this work, we address music representation learning using convolution-free transformers.
We build on top of existing spectrogram-based audio transformers such as AST and train our models on a supervised task using patchout training similar to PaSST.
In contrast to previous works, we study how specific design decisions affect downstream music tagging tasks instead of focusing on the training task.
We assess the impact of initializing the models with different pre-trained weights, using various input audio segment lengths, using learned representations from different blocks and tokens of the transformer for downstream tasks, and applying patchout at inference to speed up feature extraction.
We find that 1) initializing the model from ImageNet or AudioSet weights and using longer input segments are beneficial both for the training and downstream tasks, 2) the best representations for the considered downstream tasks are located in the middle blocks of the transformer, and 3) using patchout at inference allows faster processing than our convolutional baselines while maintaining superior performance.
The resulting models, MAEST,\footnote{
Music Audio Efficient Spectrogram Transformer.
Code for training: \url{https://github.com/palonso/MAEST}.
This model is part of Essentia models: 
\url{https://essentia.upf.edu/models.html}
} are publicly available and obtain the best performance among open models in music tagging tasks.


\end{abstract}

\section{Introduction}\label{sec:introduction}

The goal of representation learning is to develop features that are suitable for a variety of tasks, rather than being specific to the training objective. In the context of audio, these features are sometimes referred to as embeddings, and they typically have a much lower dimensionality than the original signals, making them easier to store and process. When the embeddings are well-suited to a downstream task, it is often possible to achieve good performance using shallow models that require few resources to train and run. Additionally, using a single embedding model to feed several shallow classifiers or regressors is more efficient than having individual end-to-end models, and it simplifies addressing new related tasks with minimal additional effort. As a result, embedding models are valuable for a diverse range of applications, from quick prototyping without requiring detailed knowledge of audio processing to large-scale processing of industrial audio databases.

The universal success of transformers in text~\cite{vaswani2017attention}, vision~\cite{dosovitskiy2020image}, and audio~\cite{gong2021ast} tasks motivate further research using this architecture for music representation learning.
However, most state-of-the-art (SOTA) models are based on convolutional neural networks (CNNs)~\cite{huang2020large,mccallum2022supervised,alonso2022music,huang2022mulan}.
We hypothesize that transformers are not ruling this domain yet because they require large amounts of data and computational power to overcome their convolutional counterparts, while such resources are not always available.
To address these challenges, we propose leveraging a large collection of 3.3 M tracks annotated with public-domain metadata from Discogs and using techniques to train transformers efficiently. Specifically, we focus on PaSST~\cite{koutini2021efficient}, a method that has demonstrated remarkable performance in the AudioSet~\cite{gemmeke2017audio} benchmark. This method uses patchout, a technique consisting of discarding parts of the input to regularize the training process, while also allows reducing the GPU memory and computations required for training.
In this work, we investigate the effectiveness of this technique for music representation learning, considering the impact of specific design aspects.

We focus on the impact of using different combinations of tokens from different blocks of the transformer as embeddings, starting the training from different pre-trained weights from publicly available models, using different input segment lengths, and using patchout at inference time to speed up the embedding extraction. Our experiments show that the best performance is obtained by extracting embeddings from the middle of the transformer and initializing it with weights pre-trained on other audio tasks. Contrary to previous studies based on CNNs, our transformers benefit from long input segments both in training and different downstream scenarios.
Finally, we show that, on certain patchout conditions, our transformers are able to double the inference speed of an EfficientNet-B0 baseline while producing embeddings that obtain better performance on downstream tasks. Moreover, this approach has the advantage of being entirely configurable at inference time, allowing the throughput/performance tradeoff to be adapted to the task at hand.


The remainder of this paper is structured as follows:
In Section~\ref{sec:background} we present existing works related to this study.
The experimental setup is presented in Section~\ref{sec:setup}, and the proposed experiments and results are in Section~\ref{sec:results}. 
Finally, we conclude in Section~\ref{sec:conclusions}.

\begin{figure*}
    \centering
    \includegraphics[width=\linewidth]{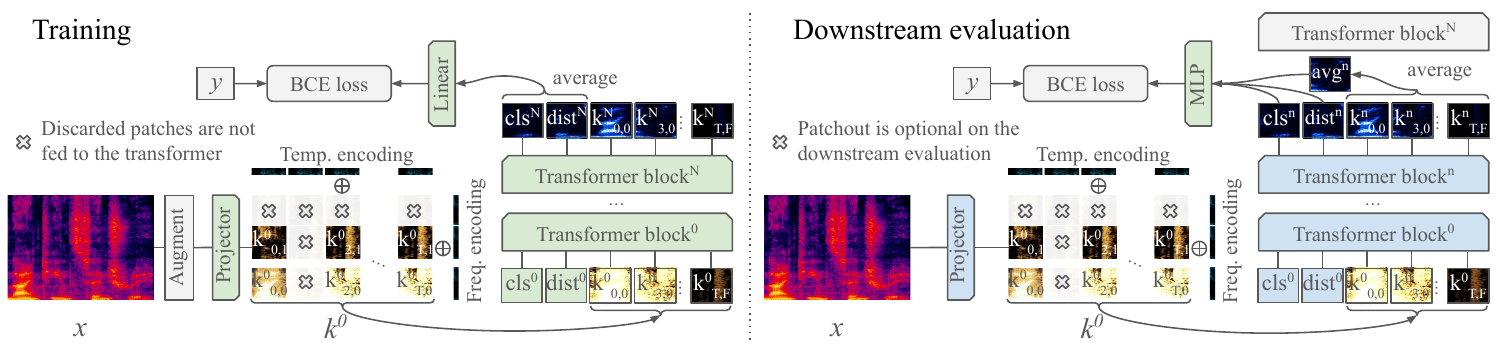}
    \caption{Illustration of our system at the training and downstream evaluation stages where $x$ is the input spectrogram, $k^0$ is the sequence of tokens after the patchout, $y$ is the target labels, and BCE is the binary cross-entropy loss.  Trainable and frozen blocks are colored green and blue respectively.}
    \label{fig:diagram}
\end{figure*}

\section{Background}\label{sec:background}
In this section, we review the literature on music representation learning to motivate the selection of our training task and discuss existing audio and music transformers and justify our architecture and training approach.
Finally, we introduce existing works on music representation learning with transformers.

\subsection{Music representation learning}

Some authors have pursued general-purpose representation models to address simultaneously speech, audio event, and music tasks, which led to the proposal of challenges such as HEAR~\cite{turian2022hear} and benchmarks such as HARES~\cite{wang2022towards}.
However, for now, there is no evidence that a single training paradigm can yield excellent performance in all the audio domains at the same time.
Alternatively, audio representations can be optimized to a single domain leveraging specific data, which tends to produce better performance.
In this sense, music-specific representation models are typically evaluated in music description in terms of genre, mood, era, rhythmic properties or arousal and valence estimation, where the annotations are generally on the track level.
Additionally, music representation models can be evaluated in more objective tasks such as tempo or key estimation, although, specific models using domain knowledge tend to be better suited for these tasks~\cite{SchreiberM18_TempoCNN_ISMIR}.

Music tagging is a multi-label classification task using a vocabulary that can combine multiple music notions (e.g., genre, moods, eras).
Some of the most successful music representation learning approaches are based on music tagging~\cite{Oord2014TransferLB,choi2017transfer,lee2018samplecnn, mccallum2022supervised}.
Other directions include training models on editorial metadata~\cite{Park2018RepresentationLO,Kim2018,Lee2019,Kim2020,huang2020large,alonso2022music,alonso23pretraining}, multi-modal correspondence~\cite{cramer2019look}, co-listening statistics~\cite{huang2020large}, contrastive supervised~\cite{Favory2020COALACA,favory21coala,Ferraro2021Enriched,huang2022mulan} and self-supervised~\cite{spijkervet2021contrastive,niizumi2021byol,yao2022contrastive,zhao2022s3t,wang2022towards} objectives, music generative models~\cite{castellon2021codified}, playlist co-occurrences~\cite{Ferraro2021Enriched, alonso23pretraining}, text~\cite{manco2021learning,huang2022mulan}, or combinations of them~\cite{huang2020large,Kim2020,castellon2021codified,Ferraro2021Enriched}.
While self-supervised approaches have been narrowing the gap with their supervised counterparts, the SOTA models 
use music tagging~\cite{huang2020large,mccallum2022supervised}, or supervised contrastive learning in a single-domain~\cite{alonso2022music} or cross-domain~\cite{huang2022mulan} settings.
Since the scope of this work is to assess the benefits of transformers, we fix our training task to music tagging for its simplicity, popularity, and empirically shown effectiveness.



\subsection{Transformers in audio classification tasks}


    
Transformers have become a popular choice for audio tasks due to their superior performance compared to their convolutional counterparts when sufficient data is available.
Lately, AudioSet, with almost 2 M audio event excerpts, has become a popular benchmark led by transformer models.
A popular approach consists of applying attention over small overlapping patches (e.g., 16 $\times$ 16) from the spectrogram using a classification objective.
The sequence of spectrogram patches is linearly projected to a 1-D space where a trainable positional encoding signal is added.
A trainable classification token is appended to the sequence of projections, and after a number of Transformer blocks it is used to solve the classification task using a linear classifier.
This idea was first introduced in the image domain by ViT~\cite{dosovitskiy2020image} and adapted to audio spectrograms in AST~\cite{gong2021ast}.
PaSST extends this approach by introducing \textit{patchout}, a technique consisting of discarding random patches from the input spectrogram at training time (see Figure~\ref{fig:diagram})~\cite{koutini2021efficient}.
This technique has two benefits.
First, by discarding input patches, the training sequence length is significantly reduced, which increases the training speed.
Second, it acts as a regularization technique that improves the robustness of the transformer.
Additionally, patchout can be combined with other training methods. 
MaskSpec is a self-supervised pre-training method based on an encoder-decoder architecture where the decoder has to reconstruct the spectrogram from a partial spectrogram altered with patchout~\cite{chong2022masked}.
Beats is a transformer trained with a supervised objective and patchout where the labels come from a codebook of randomly initialized vectors that is iteratively optimized~\cite{chen2022beats}.
While these techniques prevent the transformers from depending on initializing from weights of pre-trained models, such systems are significantly more resource-demanding.
Table~\ref{tab:audioset_transformers} compares the mentioned audio transformers in terms of GPUs used for training, training duration, and mean Average Precision (mAP) on AudioSet.
Remarkably, PaSST achieves an excellent trade-off between mAP and needed resources.
Since we aim to use transformer models that can be trained with a computational budget equivalent to SOTA CNNs (i.e., using consumer-grade GPUs), we focus on the standard patchout training with a supervised objective.

\begin{table}[]
    \centering
    \small
    \begin{tabular}{lllll}
    \toprule
    Model & Init. & GPUs & Time & mAP \\
    \midrule
    AST\cite{gong2021ast} & ViT & - & - & 45.9 \\
    PaSST\cite{koutini2021efficient} & DeiT & 2 RTX 2080ti & 24 h & 47.6 \\
    MaskSpec\cite{chong2022masked} & FS & 64 Tesla V100 & 36 h &  47.3 \\
    Beats\cite{chen2022beats} & FS & 16 & - & 48.7 \\
    \bottomrule
    \end{tabular}
    \caption{Comparison transformers from the literature in terms of initialization weights, number of GPUs used for training, training time, and mAP obtained in AudioSet.}
    \label{tab:audioset_transformers}
\end{table}


\subsection{Music representation learning with transformers}
Some works already combined music representation learning and pure-attention-based transformers.
S3T combines MoCo's momentum-based self-supervised contrastive learning with the Swin Transformer~\cite{liu2021swin} architecture to learn music representations for classification~\cite{zhao2022s3t}.
MuLan is an audio representation model trained with cross-domain contrastive learning that aligns the latent representations of associated audio and text pairs.
The authors experiment both with a ResNet50 and an AST architecture, with the former obtaining better performance in downstream music tagging tasks~\cite{huang2022mulan}.

The limited list of studies combining transformers and music representation learning motivates further research.
We propose addressing this by using a simple supervised objective and patchout.



\section{Experimental setup}\label{sec:setup}
We train our models using an in-house dataset with 3.3 M tracks mapped to the Discogs' public metadata dump.\footnote{\url{https://www.discogs.com/data/}}
The training task consists of a multi-label classification of the top 400 music styles from Discogs' taxonomy.
We compare different training configurations in several downstream tasks by training Multi-Layer Perceptrons (MLP) on representations extracted from the transformers.

\subsection{Dataset and pre-processing}
Our dataset is derived from a pool of 4 M audio tracks mapped to the release information from the Discogs website's public dump.~\footnote{In Discogs, releases include albums, EPs, compilations, etc.}
All release metadata, which can include music style tags following a pre-defined taxonomy, is submitted by the community of platform users.
\textit{Master releases} group different versions of the same release such as special editions, or remasters.
We obtain our training labels, $y$, at the master release level by first aggregating the style tags of all the associated releases and then discarding master releases with more than five style tags or without any style label among the 400 most frequent among our pool of tracks.
We keep tracks longer than 20 seconds.
Since the style annotations are done at the master release level, the resulting track annotations are expected to be noisy.
We generate validation and testing subsets with approximately 40,000 tracks and a training set with 3.3 M tracks, ensuring that every artist appears on a single split.
This pre-processing is similar to our previous work~\cite{alonso2022music}, and additional details and statistics about the resulting dataset can be found in the repository accompanying this publication.
For now on, we refer to this internal dataset as \textit{Discogs20}.

From every track, we sample 30 seconds from the center of the track and downmix it to a mono channel at 16 kHz. 
We extract 96-bands mel-spectrograms, $x$, using 32 ms windows and a hop size of 16 ms compressed with the expression $log_{10}(1 + 10000 x)$ similar to previous works in music tagging~\cite{pons2019musicnn, alonso2022music}.
The resulting representations are stored as half-precision floats (16 bits) resulting in 1.3 TB of data.
Given that our dataset is in the order of magnitude of AudioSet (1.8 M vs. 3.3 M) and presents similar label density (2.7 average labels in AudioSet and 2.1 in Discogs20), we adopt the sampling strategy used in previous works~\cite{koutini2021efficient}.
Every epoch, we take a balanced sample of 200,000 tracks without replacement using the inverse label frequencies as sample weight.
We normalize the input to the mean and standard deviation of the training set.

\subsection{Model and training}

Our transformer, \textit{MAEST}, has the same architecture as AST\cite{gong2021ast}, ViT\cite{dosovitskiy2020image}, or PassT~\cite{koutini2021efficient}, and features 12 blocks of self-attention plus a dense layer resulting in close to 87 million parameters.
We use 16 $\times$ 16 patches, $x_{t,f}$, with a stride of 10 $\times$ 10.
Similar to PaSST, we split the positional encoding into time/frequency encodings ($te_t$, $fe_f$)  and apply patchout by randomly discarding entire rows and columns from the sliced spectrogram.
The input sequence of tokens, $k^0$, is created as a linear projection of the patches plus the correspondent time/frequency encodings, $k^0_{t,f} = P(x_{t,f}) + te_t + fe_f$, where $P(\cdot)$ is a trainable linear layer.\footnote{Since the mel scale is not linear, we considered specialized projectors for each frequency patch. However, this did not improve the performance.}
$k^1$ to $k^{12}$ represent the output tokens of the respective transformer blocks.
Similar to DeiT\cite{touvron2021training} and PaSST, we extend $k^0$ with classification ($cls^0$) and distillation ($dist^0$) trainable tokens, which are initialized with the DeiT or PaSST pre-trained weights in the experiments involving these models.
\footnote{
We considered a teacher-student approach similar to DeiT
by using a pre-trained MAEST-30 to generate pseudo-labels that were targeted by the $dist^{12}$ token in the training stage.
We decided to omit the experiment details since it did not achieve a significant improvement.
}
We take the average of $cls^{12}$ and $dis^{12}$ tokens to feed a linear classifier targeting $y$.

We use the Adam Optimizer with a weight decay of $1\mathrm{e}{-4}$ and train the model for 130 epochs.
We warm up the model for 5 epochs and then keep the learning rate at $1\mathrm{e}{-4}$ until epoch 50. Then the learning rate is linearly decreased  to $1\mathrm{e}{-7}$ during 50 additional epochs.
We consider two sets of weights for inference: those from the last epoch and those obtained by taking the mean of the model's weights every 5 epochs from epoch 50 using Stochastic Weight Averaging (SWA).
We pre-compute the mel-spectrograms for efficiency, which limits the set of data augmentations we could apply. We use mixup~\cite{zhang2017mixup} with $alpha=0.3$ and SpecAugment~\cite{park2019specaugment} by masking up to 20 groups of 8 timestamps and up to 5 groups of 8 frequency bands.\footnote{We trained MAEST using 4 Nvidia 2080 RTX Ti GPUs with 12GB of RAM.
The training takes 31 hours for MAEST-5 and 48 hours for MAEST-30.}

\textbf{Initialization weights.}
Previous works showed the importance of initializing the transformer to weights pre-trained on ImageNet~\cite{gong2021ast}.
To gain further knowledge, we consider three initialization options: the DeiT B↑384 model pre-trained on ImageNet~\cite{touvron2021training}, the PaSST S S16 model pre-trained on mel-spectrograms from AudioSet, and random initialization.

\textbf{Spectrogram segment length}. We consider spectrogram segment lengths of 5 to 30 seconds resulting in the architectures MAEST-5s, MAEST-10s, MAEST-20s, and MAEST-30s.
In all cases, we take existing PaSST frequency and temporal encodings and interpolate them to the target shape as an initialization.
We use patchout discarding 3 frequency and 15 temporal patches for MAEST-5s and increase the temporal patchout proportionally for models with longer input sequences (e.g., 60 patches for MAEST-20s).

\subsection{Evaluation}
\label{sec:evaluation}
\begin{table}[]
    \centering
    \small
    \begin{tabular}{llllll}
    \toprule
        Dataset & Size & Lab. & Dur. & Av. & Split \\
    \midrule
        MTGJ-Genre & 55,215 & 87 & FT & 2.44 & split 0~\cite{bogdanov2019mtg}\\
        MTGJ-Inst & 25,135 & 40 & FT & 2.57 & split 0~\cite{bogdanov2019mtg} \\
        MTGJ-Moods & 18,486 & 56 & FT & 1.77 & split 0~\cite{bogdanov2019mtg} \\
        MTGJ-T50 & 54,380 & 50 & FT & 3.07 & split 0~\cite{bogdanov2019mtg} \\
        MTT & 25,860 & 50  & 29s & 2.70 & 12-1-3~\cite{van2014transfer} \\
        MSDs & 241,889 & 50  & 30 & 1.72 & usual~\cite{lee2018samplecnn}\\
        MSDc & 231,782 & 50  & 30 & 1.31 & CALS~\cite{Won2021SemisupervisedMT} \\
    \bottomrule
    \end{tabular}
    \caption{Automatic tagging datasets used in the downstream evaluation.
    The datasets are compared in terms of sample size, number of labels, audio duration (Full Tracks or excerpts of fixed duration), average labels per track, and the splits used in our evaluations.
    }
    \label{tab:datasets}
\end{table}

We evaluate our models in several music automatic tagging datasets covering various musical notions.
We consider the popular MagnaTagATune (MTT) and the Million Song Dataset (MSD) with the commonly used training, validation, and testing splits used in~\cite{van2014transfer} and~\cite{lee2018samplecnn} respectively.
Additionally, we report the performance of our models in the CALS split, which is an artist-filtered version of the MSD ground truth~\cite{Won2021SemisupervisedMT}.
Finally, we use the MTG-Jamendo Dataset, a dataset of Creative Commons music containing sub-taxonomies with the tags related to genre (MTGJ-Genre), moods and themes (MTGJ-Mood), and instrumentation (MTGJ-Inst), along with the top 50 tags (MTGJ-T50) in the dataset.
We use the official split 0 for all the subsets similar to previous works~\cite{won2020evaluation,manco2021learning,mccallum2022supervised}.
Table~\ref{tab:datasets} summarizes these datasets in terms of size, number of labels, audio duration, average number of labels per track, and used splits.

We evaluate our models by extracting internal representations from different blocks of the transformer and training MLP classifiers on top.
Instead of averaging the $cls^{12}$ and $dist^{12}$ tokens as done in the training stage, we consider three types of representations, $cls^{n}$, $dist^{n}$, and the average of the tokens representing the input spectrogram patches ($avg^{n}$) after $n$ transformer blocks.
Additionally, we evaluate the complementarity of these embeddings training MLP classifiers on stacks of the different tokens.
To generate the dataset of embeddings, we average the embeddings extracted from half-overlapped segments across the entire audio available for the tracks in the downstream datasets.
The same setup is used for the training, validation and testing stages.

The downstream model is an MLP with a single-hidden layer of 512 dimensions with a ReLU activation and dropout.
In the experiments described in Sections~\ref{sec:extracting_embeddings}, \ref{sec:initial_weights}, \ref{sec:input_segment}, and \ref{sec:fast_extraction}, we use a batch size of 128, drop out of 0.5 and train the model for 30 epochs.
In the downstream evaluation from Section~\ref{sec:downstream_tasks}, we perform a grid search over the following hyper-parameters for each task:
\begin{itemize}
    \item \textbf{batch size}: \{64, 128, 256\}
    \item \textbf{epochs}: \{30, 40, 50, 60, 70, 80\}
    \item \textbf{drop out}: \{0.5, 0.75\}
    \item \textbf{maximum learning rate}: \{$1\mathrm{e}{-3}$, $1\mathrm{e}{-4}$, $5\mathrm{e}{-4}$, $1\mathrm{e}{-5}$\}
\end{itemize}

The MLP is trained with the binary cross-entropy loss using the Adam optimizer with a weight decay of $1\mathrm{e}{-3}$.
The learning rate is exponentially raised to its maximum value during the first 10 epochs, kept constant for the number of epochs, and linearly reduced until reaching $1\mathrm{e}{-7}$ at the end of training.
After training, we report the performance on the testing set obtained using the weights from the epoch with the highest validation ROC-AUC.

\section{Experiments and results}\label{sec:results}
In this section, we present the conducted experiments and discuss the results.

\subsection{Extracting embeddings from the transformer}
\label{sec:extracting_embeddings}

\begin{figure}
    \centering
    \includegraphics[width=\columnwidth]{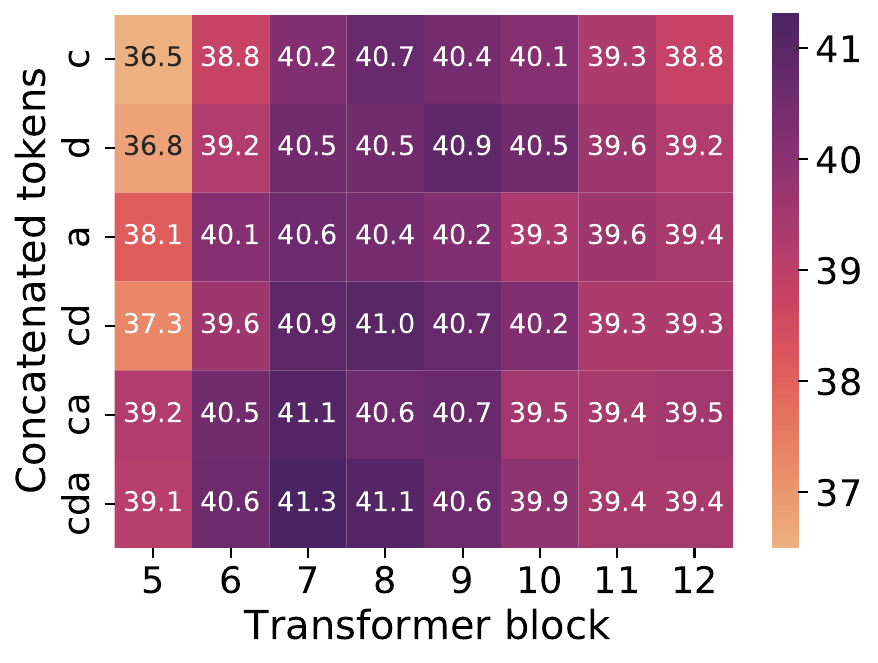}
    \caption{
    mAP scores obtained with our evaluation setup in the MTT dataset using embeddings extracted from different blocks and tokens transformer.
    We evaluate the $cls$ (c), $dist$ (d), and $avg$ (a) tokens and stacks of their combinations extracted from the transformer blocks 5 to 12.
    }
    \label{fig:heatmat}
\end{figure}

We are interested in finding the optimal representations from the transformer to be used as embeddings.
To do this, we extract representations $cls^n$, $dist^n$, and $avg^n$ from different transformer blocks $n \in [5, 12]$ .
To measure the complementarity of these features, we train MLPs fed with stacks of combinations of these representations.
In this experiment, we use MAEST-30s intialized with PaSST weights and the MTT dataset.

Figure~\ref{fig:heatmat} shows mAP scores obtained with different stacks of embeddings extracted from the different transformer blocks.
In accordance with previous studies~\cite{castellon2021codified}, we find that the embeddings with the best performance are found in the middle blocks of the transformer.
This contrasts with the typical behavior of CNNs, where the best features are normally towards the last layers of the model, especially, when the downstream task is well aligned with the training task.
Also, concatenating the features benefits the performance.
In the remaining experiments, we fix our embedding to the stack
($cls^7$, $dist^7$, $avg^7$).

\subsection{Impact of the initial weights}
\label{sec:initial_weights}

Due to the lack of inductive biases present in architectures such as CNNs, transformers are heavily dependent on pre-training.
Because of this, many audio transformers are initialized with weights transferred from image tasks~\cite{gong2021ast, koutini2021efficient}.
We evaluate the impact of initializing our models from the weights of DeiT~\cite{touvron2021training} (image input), the best single PaSST model~\cite{koutini2021efficient} (mel-spectrogram input), and random initialization.
In this experiment, we use MAEST-10s and its version with SWA weights, MAEST-10s-swa.
Although our main focus is to evaluate MAEST on public downstream datasets, we also report their performance on the training task to provide additional insights.

Table~\ref{tab:inital_weights} shows the performance in both, the training (Discogs20), and a downstream (MTT) task.
In both cases, the scores are higher when the training is started from pre-trained weights.
Since the PaSST weights result in slightly higher performance, we use this initialization for the remaining of this work.
Regarding the SWA, we observe a positive effect on the training task when the model is initialized with pre-trained weights.
However, we do not observe improvements in the downstream task.

\begin{table}[t]
    \small
    \centering
    \begin{tabular}{lccc}
    \toprule
    Model & RW & DeiT & PaSST \\
    \midrule
    \multicolumn{4}{l}{\scriptsize \textit{Pre-training task: Discogs20}} \\
    MAEST-10s & 20.5 & 22.7 & 22.8 \\
    MAEST-10s-swa & 20.1 & 23.2 & 23.5 \\
    \midrule
    \multicolumn{4}{l}{\scriptsize \textit{Downstream task: MTT}} \\
    MAEST-10s & 38.7 & 40.4 & 41.1 \\
    MAEST-10s-swa & 39.0 & 40.2 & 41.0 \\
    
    \bottomrule
    \end{tabular}
    \caption{ 
    mAP scores obtained in the training and downstream tasks using different initializations.
    We considered Random Weights, and pre-trained weights from DeiT and PaSST.
    }
    \label{tab:inital_weights}
\end{table}

\subsection{Effect of the input segment length}
\label{sec:input_segment}

\begin{table}[b]
    \centering
    \small
    \begin{tabular}{lcccc}
         \toprule
         Model & 5s & 10s & 20s & 30s \\
        \midrule
        \multicolumn{5}{l}{\scriptsize \textit{Pre-training task: Discogs20}} \\
        MAEST-\textit{T} & 21.1 & 22.8 & 24.8 & 26.1\\
        MAEST-\textit{T}-swa & 21.3 & 23.5 & 25.8 & 27.0 \\
        \midrule
        \multicolumn{5}{l}{\scriptsize \textit{Downstream task: MTT}} \\
         MAEST-\textit{T} & 40.8 & 41.1 & 41.2 & 41.7 \\
         MAEST-\textit{T}-swa & 40.9 & 41.0 & 41.2 & 41.5 \\
         \bottomrule
    \end{tabular}
    \caption{
    mAP scores obtained in the training and downstream tasks using different spectrogram segment lengths.
    \textit{T} represents the spectrogram segment length.
    }
    \label{tab:segment_length}
\end{table}

We train MAEST using input segment lengths ranging from 5 to 30 seconds.
In our experiments, we keep the frequency patchout constant and proportionally increase the temporal patchout.
For our models with segment lengths of 5, 10, 20, and 30 seconds we discard 15, 30, 60, and 90 temporal patches respectively.

Table~\ref{tab:segment_length} shows the performance of the MAEST models with respect to their input spectrogram segment length in terms of mAP both in the training (Discogs20) and a downstream (MTT) evaluation.  
While music tagging CNNs tend to reach their peak of performance with receptive fields of 3 to 5 seconds\cite{choi2017transfer}, 
attention-based systems have shown the capability to take advantage of longer temporal contexts~\cite{Won2021SemisupervisedMT}.
Our models are consistent with this trend, reaching their best performance when trained on segments of 30 seconds.
Although even longer segments could be beneficial, we could not use them while keeping the same model size due to GPU memory limitations.

\subsection{Performance in downstream tasks}
\label{sec:downstream_tasks}

\begin{table*}
\centering
\small
\begin{tabular}{lrrrrrrrrrrrrrr}
\toprule
  & \multicolumn{2}{c}{MTGJ-Genre} & \multicolumn{2}{c}{MTGJ-Inst} & \multicolumn{2}{c}{MTGJ-Mood} & \multicolumn{2}{c}{MTGJ-T50} & \multicolumn{2}{c}{MTAT} & \multicolumn{2}{c}{MSDs} & \multicolumn{2}{c}{MSDc} \\
 & ROC & mAP & ROC & mAP & ROC & mAP & ROC & mAP & ROC & mAP & ROC & mAP  & ROC & mAP \\
\midrule
\multicolumn{15}{l}{\scriptsize \textit{State of the art}} \\
\multirow{2}{*}{Fully-trained}      & - & - & - & - & 77.8 & 15.6 & 83.2 & 29.8 & 90.69 & 38.44  & 92.2 & 38.9 & 89.7  & 34.8 \\
      \scriptsize & - & - & - & - & \textit{\cite{knox2020mediaeval}} & \textit{\cite{knox2020mediaeval}} & \textit{\cite{pons2019musicnn}} & \textit{\cite{pons2019musicnn}} & \textit{\cite{won2020evaluation}} & \textit{\cite{won2020evaluation}} & \textit{\cite{Won2021SemisupervisedMT}} & \textit{\cite{Won2021SemisupervisedMT}} & \textit{\cite{Won2021SemisupervisedMT}} & \textit{\cite{Won2021SemisupervisedMT}} \\
\multicolumn{15}{l}{\tiny \textit{ }} \\

\multirow{2}{*}{Embeddings} & 87.7 & 19.9 & 77.6 & 19.8 & 78.6 & 16.1 & 84.3 & 32.1 & 92.7 & 41.4 & - & - & 90.3 & 36.3 \\
  & \textit{\cite{alonso2022music}} & \textit{\cite{alonso2022music}} & \textit{\cite{alonso2022music}} & \textit{\cite{alonso2022music}} & \textit{\cite{mccallum2022supervised}} $^\dagger$ & \textit{\cite{mccallum2022supervised}} $^\dagger$ & \textit{\cite{mccallum2022supervised}} $^\dagger$ & \textit{\cite{mccallum2022supervised}} $^\dagger$ & \textit{\cite{huang2022mulan}} $^\dagger$ & \textit{\cite{mccallum2022supervised}} $^\dagger$ & - & - & \textit{\cite{mccallum2022supervised}} $^\dagger$ & \textit{\cite{mccallum2022supervised}} $^\dagger$ \\

\midrule
\multicolumn{15}{l}{\scriptsize \textit{Baseline}} \\
EffNet-B0 & 87.7 & 19.9 & 77.6 & 19.8 & 75.6 & 13.6 & 83.1 & 29.7 & 90.2 & 37.4 & 90.4 & 32.8 & 88.9 & 32.8 \\

\midrule
\multicolumn{15}{l}{\scriptsize \textit{Our models}} \\
MAEST-10s & \cellcolor[gray]{0.9} 88.1 & \cellcolor[gray]{0.9} 21.1 & \cellcolor[gray]{0.9} 79.7 & \cellcolor[gray]{0.9} 22.4 & \cellcolor[gray]{0.9} 77.9 & 15.1 &  \cellcolor[gray]{0.9} 84.0 & \cellcolor[gray]{0.9} 31.3 & \cellcolor[gray]{0.9} 91.8 & \cellcolor[gray]{0.9} 41.0 & 91.5 & 36.9 & 88.9 & 32.7 \\
MAEST-20s & \cellcolor[gray]{0.9} 88.1 & \cellcolor[gray]{0.9} 21.4 & \cellcolor[gray]{0.9} 79.9 & \cellcolor[gray]{0.9} 22.6 & \cellcolor[gray]{0.9} 77.9 & 15.2 & \cellcolor[gray]{0.9} \textbf{84.1} & \cellcolor[gray]{0.9} \textbf{31.5} & \cellcolor[gray]{0.9} 91.8 & \cellcolor[gray]{0.9} 41.0 & 92.1 & \cellcolor[gray]{0.9} 39.2 & 89.5 & 34.5 \\
MAEST-30s & \cellcolor[gray]{0.9} \textbf{88.2} & \cellcolor[gray]{0.9} \textbf{21.6} & \cellcolor[gray]{0.9} \textbf{80.0} & \cellcolor[gray]{0.9} \textbf{22.9} & \cellcolor[gray]{0.9} \textbf{78.1} & \textbf{15.4} & \cellcolor[gray]{0.9} 84.0 & \cellcolor[gray]{0.9} \textbf{31.5} & \cellcolor[gray]{0.9} \textbf{92.0} & \cellcolor[gray]{0.9} \textbf{41.9} & \cellcolor[gray]{0.9} \textbf{92.4} & \cellcolor[gray]{0.9} \textbf{40.7} & \cellcolor[gray]{0.9} \textbf{89.8} & \cellcolor[gray]{0.9} \textbf{35.4} \\


\bottomrule

\label{tab:benchmark}

\end{tabular}

\caption{
ROC-AUC and mAP scores obtained in the downstream tasks.
Our baseline consists of an EffNet-B0 architecture trained in Discogs20.
Additionally, we report the SOTA results distinguishing models with all parameters trained in the downstream tasks (fully trained) and models evaluated with shallow classifiers.
For every task, we mark in bold the best score obtained by a MAEST model and highlight in grey models achieving better performance than the best open alternative. $^\dagger$ Models not publicly available.
}

\end{table*}

Considering our previous findings, we extend the evaluation of MAEST to a number of downstream datasets. 
We evaluate MAEST-10s, MAEST-20s, MAEST-30s, 
and a baseline consisting of embeddings from the penultimate layer of an EfficientNet-B0 (EffNet-B0) architecture~\cite{tan2019efficientnet} trained in the same 400 music style tags from Discogs20 following previous work~\cite{alonso2022music}.
Additionally, we report the performance of SOTA models from the literature considering approaches fully trained in the downstream tasks and based on embeddings plus shallow classifiers.

Table~\ref{sec:results} shows the results of the different models in terms of ROC-AUC and mAP.
We observe that all the MAEST models outperform the baseline in all tasks, confirming the superiority of the proposed approach.
Additionally, we achieve a new SOTA for the MTGJ-Genre, MTGJ-Inst, and MSDc datasets, although other models remain superior in the rest of the datasets.
Specifically, MuLan~\cite{huang2022mulan} obtains higher mAP in MTT, probably because it is trained on a much larger corpus of 40 M tracks.
In MTGJ-Moods, MTGJ-T50, MTT, and MSDs, Musicset-Sup, a model trained on a curated dataset of 1.8 M expert annotations, remains superior~\cite{mccallum2022supervised}.
In both cases, the advantage is likely due to the superiority of the training task. 
Notably, none of these models is public, which makes MAEST the best open music embedding extractor available.


\subsection{Faster feature extraction with inference patchout}

Inferring with transformers is typically more computationally expensive than with CNNs.
To speed up our models, we consider using two types of patchout at inference time:
Time-wise, we keep one out of $T$ spectrogram patches.
Frequency-wise, we discard specific rows of patches.
We experiment with temporal patchout using $T \in [2, 3, 5, 10 ]$ and frequency patchout of 3 and 4 patches corresponding to the first and the two last blocks, and the two first and two last blocks respectively.
The embeddings obtained under different patchout settings are compared in the training and a downstream task following our downstream evaluation approach on the MTT dataset.

\begin{figure}
    \centering
    \includegraphics[width=\linewidth]{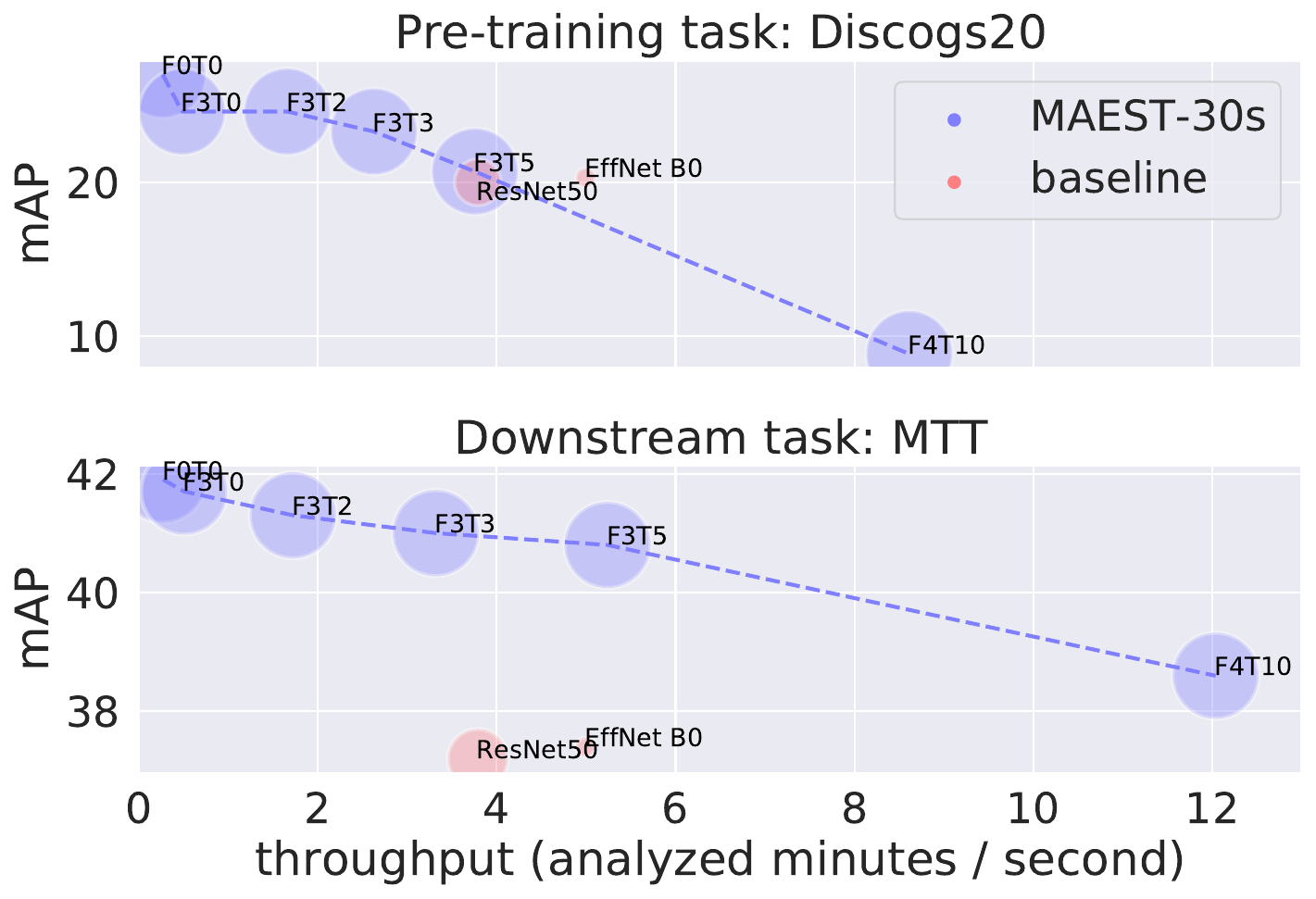}
    \caption{mAP scores against throughput for MAEST-30s under different amounts of frequency (F) and time (T) patchout.
    The radius is proportional to the parameter count and the inference is performed on the CPU.
    }
    \label{fig:efficiency}
\end{figure}

Figure~\ref{fig:efficiency} shows the mAP scores on the training and downstream tasks under different patchout settings. 
In the downstream task, even under strong patchout settings, MAEST-30s overcomes the throughput of standard CNN architectures by two to three times while keeping higher mAP.
On the training task, this technique is not so effective because the classifier is frozen and cannot adapt to the effects of patchout, and also it operates on tokens from the last block, which requires more computations.

\label{sec:fast_extraction}

\vspace{-8pt}
\section{Conclusion}\label{sec:conclusions}
\vspace{-5pt}
In this work, we demonstrate the benefits of pure-attention-based transformers for music representation learning
and study how different design decisions affect the downstream performance.
Our experiments show that the best embeddings come from a stack of features from the middle blocks of the transformer, initializing from weights pre-trained in audio event recognition provides the best performance, and that longer input segments correlate with better results.
We evaluate our models in six popular music tagging datasets, and experiment with patchout at inference time, finding that it allows speeding up significantly the transformer while producing embeddings with better performance/speed trade-offs than our convolutional baselines.
Finally, we present MAEST, a family of transformers for music style tagging and embedding extraction, which are publicly available and achieve SOTA performance among currently available music representation models.

In future work, we will combine our architecture with additional training objectives combining supervised and self-supervised paradigms.
Additionally, we will experiment with longer input segments and teacher-student setups suitable for noisy datasets such as ours.

\section{Acknowledgements}
This work has been supported by the Musical AI project - PID2019-111403GB-I00/AEI/10.13039/501100011033, funded by the Spanish Ministerio de Ciencia e Innovación and the Agencia Estatal de Investigación.
\bibliography{refs}

%
%
%
%
%

\end{document}